\documentclass[10pt,journal,compsoc]{IEEEtran}
\ifCLASSOPTIONcompsoc
  \usepackage[nocompress]{cite}
\else
  \usepackage{cite}
\fi
\ifCLASSINFOpdf
  \usepackage[pdftex]{graphicx}
  \DeclareGraphicsExtensions{.pdf,.jpeg,.png}
\else
  \usepackage[dvips]{graphicx}
  \DeclareGraphicsExtensions{.eps}
\fi

\usepackage{authblk}
\usepackage[utf8]{inputenc}
\usepackage[T1]{fontenc}
\usepackage[colorlinks,linkcolor=blue,citecolor=blue]{hyperref}
\usepackage[dvipsnames]{xcolor}
\usepackage{booktabs}
\usepackage{multirow}
\usepackage[framemethod=tikz]{mdframed}
\usepackage{balance}
\usepackage{algorithm}
\usepackage[inline]{enumitem}
\usepackage{algorithmic}
\usepackage{tabularx}
\newcolumntype{R}{>{\raggedleft\arraybackslash}X}

\usepackage{xspace}
\usepackage{pgfplots}
\usepackage[normalem]{ulem}
\pgfplotsset{compat=1.15}
\usepackage{subfig}
\definecolor{light-gray}{gray}{0.95}
\definecolor{light-blue}{RGB}{231,240,248}
\definecolor{green-highlight}{RGB}{172,229,172}

\usetikzlibrary{patterns}

\usepackage{listings}
\usepackage{soul}

\lstdefinelanguage{Dockerfile}
{
  morekeywords={FROM, RUN, CMD, LABEL, MAINTAINER, EXPOSE, ENV, ADD, COPY,
    ENTRYPOINT, VOLUME, USER, WORKDIR, ARG, ONBUILD, STOPSIGNAL, HEALTHCHECK,
    SHELL},
  morecomment=[l]{\#},
  morestring=[b]"
}

\DeclareRobustCommand{\hlgreen}[1]{{\sethlcolor{green-highlight}\hl{#1}}}

\usepackage{amsthm}
\theoremstyle{definition}

\usepackage[framemethod=tikz]{mdframed}
\mdfdefinestyle{mpdframe}{
    frametitlebackgroundcolor   =black!15,
    frametitlerule              =true,
    roundcorner                 =1pt,
    middlelinewidth             =1pt,
    innermargin                 =0.1cm,
    outermargin                 =0.1cm,
    innerleftmargin             =0.1cm,
    innerrightmargin            =0.1cm,
    innertopmargin              =0.1cm,
    innerbottommargin           =0.1cm
}

\newcommand{\reviseadd}[1]{#1}

\begin{document}

\title{Automatic Observability \\ for Dockerized Java Applications}
\author[1]{Long Zhang}
\author[1]{Deepika Tiwari}
\author[2]{Brice Morin}
\author[1]{Benoit Baudry}
\author[1]{Martin Monperrus}
\affil[1]{KTH Royal Institute of Technology, Sweden}
\affil[2]{Tellu, Norway}
\date{} 

\newcommand\nbProjects{87\xspace} 
\newcommand\nbApplicationDockerFiles{170\xspace} 
\newcommand\nbBaseImages{86\xspace} 

\IEEEtitleabstractindextext{%
\begin{abstract}
Docker is a virtualization technique heavily used in the industry to build cloud-based systems. In the context of Docker, a system is said to be observable if engineers can get accurate information about its running state in production.
In this paper, we present a novel approach, called POBS, to automatically improve the observability of Dockerized Java applications. POBS is based on automated transformations of Docker configuration files. Our approach injects additional modules in the production application, in order to provide better observability.
We evaluate POBS by applying it on open-source Java applications which are containerized with Docker. Our key result is that 148/\nbApplicationDockerFiles (87\%) of Docker Java containers can be automatically augmented with better observability.
\end{abstract}

}

\maketitle

\IEEEraisesectionheading{\section{Introduction}}

\IEEEPARstart{D}{ocker} is a virtualization technique heavily used in the industry to build cloud-based systems \cite{7922500}, and in software engineering research to improve reproducibility \cite{7883438,Horton:ICSE2019:DockerizeMe}. The main reason behind its success is that it encapsulates dependencies, it simplifies deployment, and its stateless nature increases reliability. As such, Docker is an important technology behind the so-called `continuous deployment' advocated by the DevOps movement \cite{Henkel:DevOpsArtifactsForDocker}. Docker is a layer directly on top of the operating system, and can be used with many programming languages and runtimes, including Java \cite{JavaInsideDocker}. Packaging Java applications inside Docker containers is currently one of the major trends in enterprise systems \cite{DevOpsAndDockerTrends2019}.

Observability is a major challenge for cloud-based systems \cite{beyer2016site}. The observability problem means that it is hard for engineers to get accurate information about the running state in production. In the old days of monolithic applications, engineers only had to look at one single process and its log file to monitor a service. Now, imagine dozens of Docker nodes, independently built by different teams and asynchronously deployed by an orchestration system such as Kubernetes: it becomes really hard to understand what happens in production \cite{beyer2016site}. To identify the root-cause of bugs and performance bottlenecks, this lack of observability can become a show-stopper \cite{observability_beyond_logging}. 
In this paper, we improve the engineering of Java applications that are packaged and deployed with Docker.

We propose a technique, called `POBS' (standing for imProved OBServability), to statically analyze and transform Docker configuration files of Java applications in order to inject observability capabilities. Our key insight is to leverage the declarative nature of Docker configuration files. 
POBS consists of parsing Docker configuration files and injecting tailored Docker configuration directives. Those additional directives start different modules that improve observability.
Our approach is fully automated, it addresses the typical case in industry of having many dockerized Java applications, automatically packaged and continuously deployed with Docker.
To our knowledge, POBS is the first-ever technique to automatically improve the observability of Docker applications in cloud-based systems with automated Docker file transformation.
For example, POBS allows developers to observe the JVM memory or CPU usage of their application with minimal effort: a single line change in the Docker configuration.
We perform a large scale empirical evaluation of POBS, by applying it to real-world open-source dockerized Java applications. We first curate a set of \nbApplicationDockerFiles Java application Docker configuration files. Next, we apply our technique to all of them. Our key result is that POBS automatically adds observability capabilities to 148/\nbApplicationDockerFiles applications (87\%), demonstrating the wide applicability of our approach.

To sum up, our contributions are:

\begin{itemize}

\item an original empirical study of the usage of Docker for Java applications in open-source projects, showing the extreme popularity of a handful of Docker base images: the 25 most popular base images cover $40.5\%$ of applications. To our knowledge, this is the first study of this kind in the literature.

\item a novel approach, called POBS, for automatically adding observability in Dockerized Java applications, based on automated transformation of Docker configuration files. The concept of automatically transforming Docker files is highly novel, and to nurture this research direction, we provide the community with a publicly-available prototype implementation for future research.

\item an evaluation of POBS on \nbApplicationDockerFiles real Dockerized Java applications found on GitHub, showing that 148/\nbApplicationDockerFiles (87\%) of these applications can be augmented with better observability, in an automated manner. Our idea of transforming Docker configuration files is applicable in practice and relevant for industry.

\item an original qualitative case study on using improved observability in Dockerized Java applications for assessing resilience with low overhead.

\end{itemize}

\section{Background}

\subsection{Docker}\label{sec:background-docker}
In modern software development pipelines, containerization tools such as Docker have become increasingly popular~\footnote{As of this writing, Docker has more than 32,000 stars on GitHub and nearly 105 billion container downloads} as they allow applications to be packaged along with their dependencies and configuration as one standalone entity. These entities, called containers in Docker, can be shipped and deployed to a host and run as a process along with other containers, sharing the same underlying operating system kernel but with their own private filesystem. Compared with virtualization based on virtual machines, containerization is a more lightweight and scalable approach.

\begin{figure}[h]
\centering
\includegraphics[width=\columnwidth]{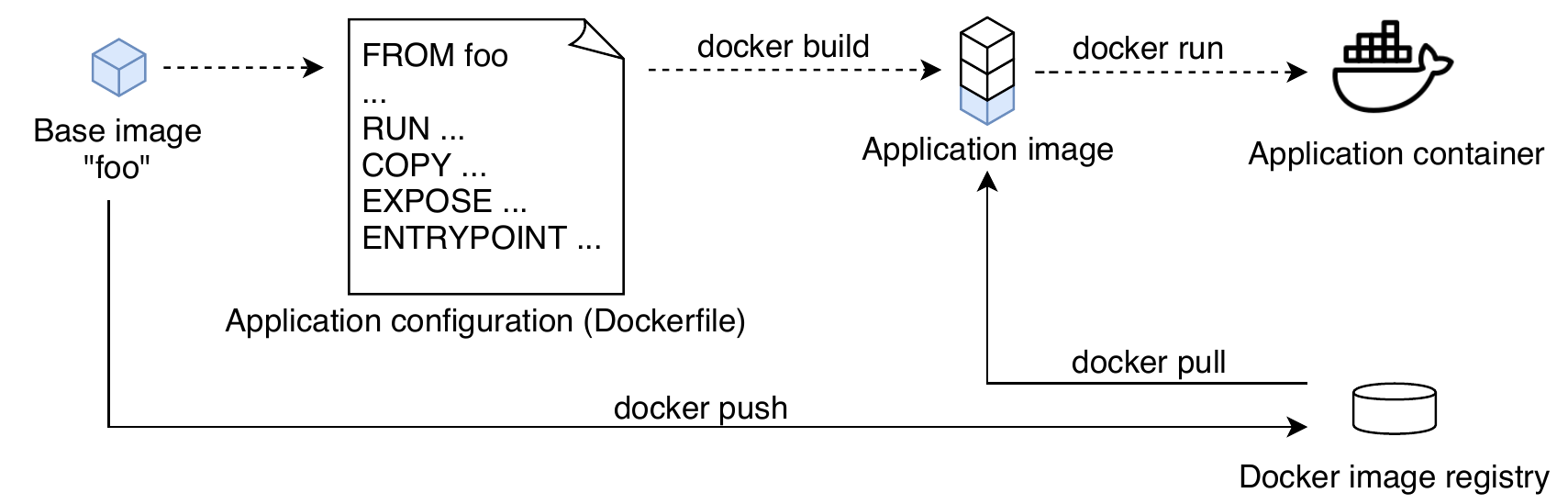}
\caption{Lifecycle of a Docker application}
\label{fig:docker}
\centering
\end{figure}

\autoref{fig:docker} summarizes a typical Docker lifecycle. A developer first defines a \emph{Dockerfile}. This file includes a mandatory \texttt{FROM} instruction, which defines a \emph{base image}. The file also specifies all the instructions needed to build~\footnote{\url{https://docs.docker.com/engine/reference/builder}} a \emph{Docker image} for an application. Each instruction in a Dockerfile (\texttt{RUN}, \texttt{COPY}, \texttt{EXPOSE}, \texttt{ENTRYPOINT}, ...) forms a new layer atop the previous one.

When a Dockerfile such as the one in \autoref{lst:dockerfile} is defined, it is possible to build a \emph{Docker image}. This build process consists of following each of the instructions in the Dockerfile, some of which may import resources from the host. The Docker image can then be shipped and instantiated as a \emph{container}. The entire infrastructure for an application can thus be determined by its Docker image. Developers can make Docker images available on \emph{registries}, the largest one being Docker Hub.

A base image is the foundation on which developers build their application Docker images. Given the critical role of base images, Docker, Inc. maintains a curated set of official images~\footnote{\url{https://hub.docker.com/search/?q=&type=image&image_filter=official}} for essential operating systems, programming languages, and database runtime environments. Besides these official images, any organization can publish images of its applications to Docker Hub, such as \texttt{anapsix/alpine-java}, \texttt{azul/zulu-openjdk}, or \texttt{cloudunit/base-jessie} to name a few. Both official images and images published by organizations are versioned and can be used as base images. 

\begin{lstlisting}[caption=A basic application Dockerfile, label={lst:dockerfile}]
FROM openjdk:8-jdk
RUN install-dep.sh
COPY my-app.jar /
...
EXPOSE 8080
ENTRYPOINT ["java", "-jar", "my-app.jar"]
\end{lstlisting}

\subsection{Observability for Resilience Engineering}\label{sec:observability-for-re}

Observability \cite{MilesRuss2019ChaosObservability} is the ability to collect information about the internal state of a software system based on its external behavior. Observability is essential for root cause analysis of problems and for resilience engineering in general. The more information engineers get, the more accurately can they deduce the relationship between a failure and its impact on the system and the end-user. The observability of a software system can be described at different levels: 1) on the operating system level, where metrics like CPU, memory usage, and I/O are considered, 2) on the service level, where the functionality of a single machine or service is monitored, e.g. does the service reply to a specific request (health-check), and 3) on the application level, where multiple nodes are monitored, e.g. with distributed tracing.

In this paper, we focus on resilience engineering, adopting the definition of Trivedi \cite{trivedi2009resilience}, who states resilience as ``the persistence of service delivery that can justifiably be trusted, when facing changes.'' For resilience engineering, developers apply fault injection techniques to actively trigger the error-handling logic. For example, developers can change bits in memory to evaluate how an operating system tolerates memory errors \cite{han1993doctor}.

Support for deep observability is a costly endeavor for developers.
To gain observability, developers need to configure and interact with different monitoring and tracing frameworks, in source code and/or in the runtime environment \cite{884757}. It is even worse for fault injection: developers who wish to experiment with fault injection need to incorporate new tools into their system, which is a tedious and error-prone task \cite{6866200}.
This is where POBS makes a contribution, it enhances the observability of Java systems by automatically adding observation capabilities to applications at a minimal cost.

\section{Empirical Study of Dockerized Java Applications}\label{sec:empirical-study}
This section describes an original empirical analysis of Dockerized Java applications. 
The goal of this study is to analyze the usage of base images in open-source Java projects that use Docker.
To our knowledge, ours is the first-ever study of this kind. By analyzing how popular open-source projects use Docker, we extract unique insights about this active software ecosystem much used in industry.

\subsection{Methodology}

First, we have queried GitHub to collect the top $1000$ repositories, sorted by the number of stars, that fulfill the following criteria: 1) Java is their primary programming language, 2) they mention ``Docker'' in their README. This latter criterion indicates that they are likely to have Dockerfiles that build images for their application.

Second, in order to extract data from the Dockerfiles, if any, we clone those repositories. If there exist releases associated with a given repository, we checkout the code corresponding to the latest release version. If no release exists, we checkout the default branch of the repository at the latest commit. This results in $434$ repositories being cloned at the commit corresponding to the latest release, and $566$ repositories being cloned at the latest commit on the default branch, as of July 09, 2020.

\subsection{Results}\label{sec:empirical-study-results}
Of the $1000$ repositories that were cloned, we find that $420$ do not have a Dockerfile. A manual investigation reveals that there are repositories that mention Docker in the README but do not use it, there are a number of reasons for this: they may be a tutorial, they may provide support for Docker with manual actions from users, or they may keep the Docker configuration in a separate repository.

The remaining $580$ repositories have at least one Dockerfile, with the total number of Dockerfiles being $2071$. On analyzing these repositories, we observe that $314/580$ ($51.1\%$) of them have exactly one Dockerfile. Five repositories have more than $40$ Dockerfiles, with the maximum number of Dockerfiles being $99$. This happens if an application is large and has many services or modules all packaged as separate Docker containers.

Next, we measure the frequency of base images in the Dockerfiles in our dataset. For this, we collect their names from the \texttt{FROM} instruction. We extract $2295$ \texttt{FROM} instructions and the corresponding base images from our dataset, of which $692$ images are unique.  

\begin{figure}
\centering
\includegraphics[width=\columnwidth]{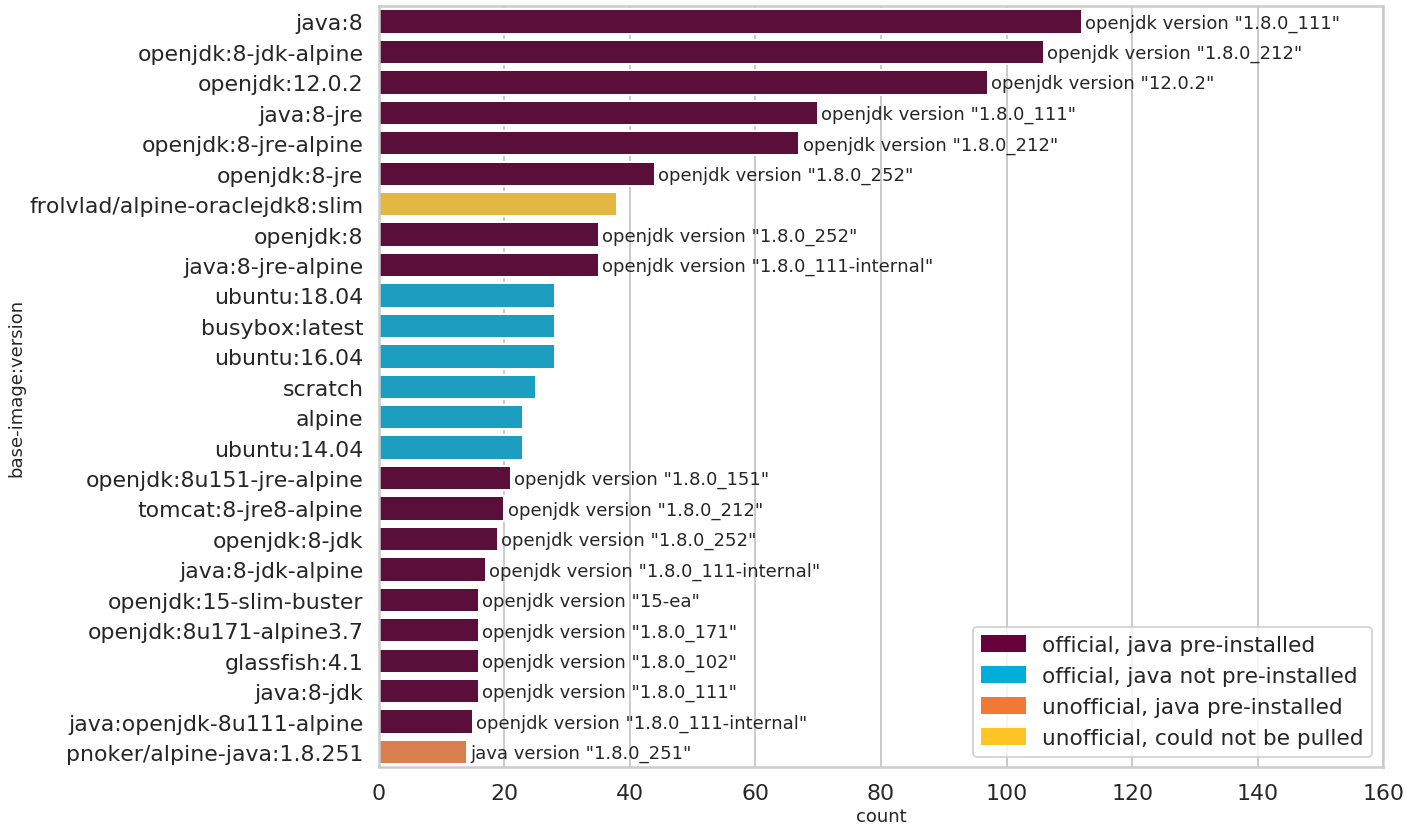}
\caption{The 25 most popular base images across 2071 Dockerfiles in the dataset. The Java version that comes pre-installed with the image, if any, is indicated. Official vendor images are represented as purple or blue bars.}
\label{fig:base-images}
\centering
\end{figure}

\autoref{fig:base-images} shows the $25$ most frequently occurring base images. Axis Y depicts a base image and Axis X depicts the number of its occurrences in our dataset. The most popular base image is \texttt{java:8} meaning the version 8 of base image \texttt{java}. This confirms that our selection criteria work as expected.
We observe that the popularity is very skewed: We observe at least one occurrence of one of these 25 images in $929/2295$ ($40.5\%$) of all base images in our dataset, indicating a skewed distribution. 
We find that $23/25$ ($92\%$) of these are listed as official images on Docker Hub, showing that developers prefer to use official images as they are well-documented and intensively reviewed for security vulnerabilities. Those official images are represented with purple and blue bars in \autoref{fig:base-images}.
Next, our analysis reveals that $18/25$ ($72\%$) of the most popular base images come with some version of Java pre-installed, suggesting that developers prefer to use base images that are already tailored for Java applications. In \autoref{fig:base-images}, we show the pre-installed Java version on the right side of the bars, clearly showing the dominance of openjdk, likely because of its friendly redistribution license~\footnote{Note that pull access was denied for the image \texttt{frolvlad/alpine-oraclejdk8:slim}, preventing us from determining the Java version, see  \url{https://github.com/Docker-Hub-frolvlad/docker-alpine-java}}. 

To sum up, for Java applications, the majority of developers rely on a small number of base images, with Java pre-installed. This confirms the feasibility of our project: by manipulating base images, we can augment a large number of Dockerized Java applications.

\subsection{Curated Dataset of Java Application Dockerfiles}\label{sec:evaluation-dataset}

We leverage on this empirical study to build a dataset usable for evaluating our contribution on engineering Java application Dockerfiles. In this section, we report on our effort to curate a set of Java application Dockerfiles that fits this need.

The first prerequisite for evaluating POBS on real-world application Dockerfiles is that we are able to build them in our lab environment without any modifications, using a uniform build command: \texttt{docker build -t TAG\_NAME -f DOCKERFILE}. This command means building an image with tag name TAG\_NAME according to the instructions mentioned in the file DOCKERFILE. The second prerequisite is to have Dockerfiles used to execute Java applications (as opposed to Dockerfiles used for testing, such as a Dockerfile that sets up a MySQL server for unit testing). Hence, we exclude Dockerfiles that are not meant to run a Java process by checking the presence of Java processes in the initialized container. \reviseadd{The last prerequisite is that these Dockerfiles  belong to real Java projects instead of simple code examples or tutorials. This requirement is manually checked by inspecting each project's GitHub repository before adding it into the dataset.}

Of the $2071$ Dockerfiles collected in \autoref{sec:empirical-study-results}, we observe that  $810$ ($39\%$) are buildable with a default \texttt{docker build} command. The most common failure, which occurs in $805/2071$ of failed cases, is ``no such file or directory''. This error means that instructions like \texttt{COPY} or \texttt{ADD} fail to find the target file. This usually happens because some of the target files need to be built beforehand. Another common failure is ``returned a non-zero code'', which occurs in $198/2071$ failing cases. The reasons for this failure are numerous, with no dominant cause. The rest of the failures are specific to the Dockerfiles, such as ``pull access denied'', ``404 not found'', etc.

\begin{figure}
    \centering
    \subfloat{\includegraphics[width=\columnwidth]{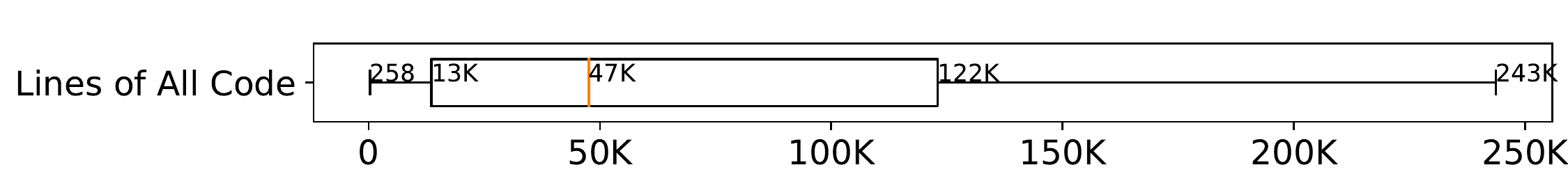}}\\
    \subfloat{\includegraphics[width=\columnwidth]{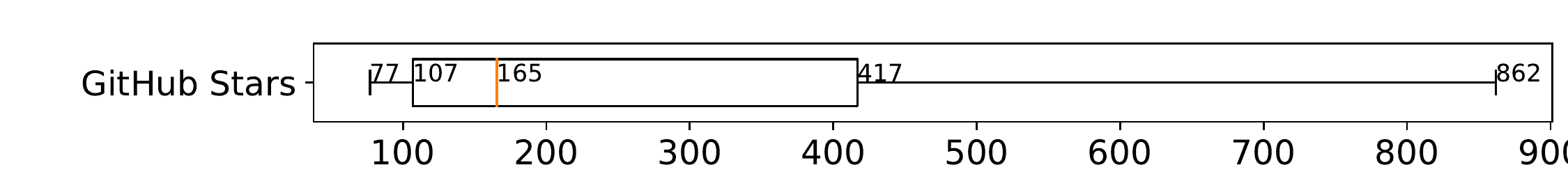}}\\
    \subfloat{\includegraphics[width=\columnwidth]{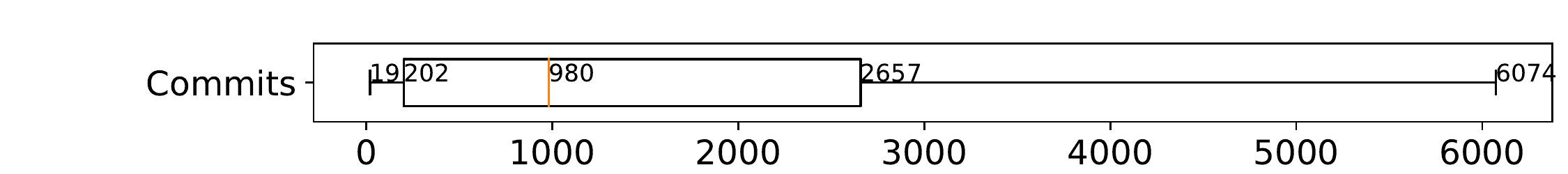}}\\
    \subfloat{\includegraphics[width=\columnwidth]{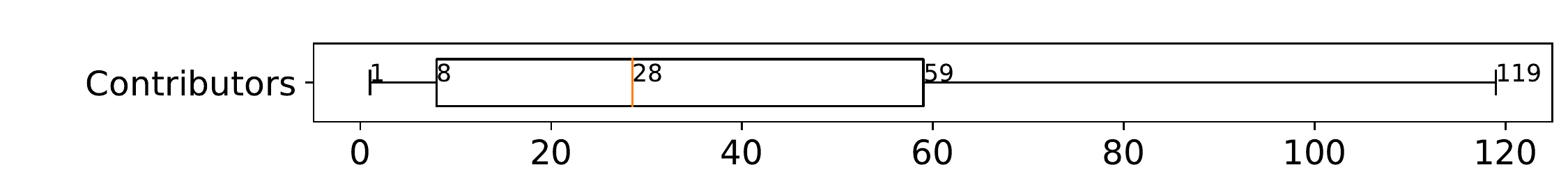}}
    \caption{Characteristics of the \nbProjects curated projects }\label{fig:covered-projects}
\end{figure}

\textbf{The final curated dataset contains \nbApplicationDockerFiles buildable and runnable Dockerfiles.}
They are spread over \nbProjects open-source Java projects which use Docker. \autoref{fig:covered-projects} describes the complexity, popularity, and maturity of these projects, with respect to the number of lines of source code, GitHub stars, the number of commits, and the number of contributors. \reviseadd{The box plots represent, from left to right, the values for the minimum, first quartile, median, third quartile, and maximum. Note that we do not include outliers in these plots.} The metadata for these Dockerfiles - their base images, arguments, commands, and entrypoints - are consolidated and made available at \url{https://github.com/KTH/royal-chaos/tree/master/pobs}.

\section{Technical Approach}
\label{sec:technical-contribution}
The goal of POBS is to improve the fine-grained observability of Dockerized Java applications in an automated manner. Once these applications provide sufficient observability, developers are able to evaluate their resilience using fault injection experiments. To this end, POBS automatically weaves in the observability and fault injection capabilities in their Docker images. In this section, we present the POBS architecture, the key design choices, and the implementation decisions made in developing our prototype.

\subsection{Working Example}
Let us recall the example of the Dockerfile in \autoref{lst:dockerfile}, where developers define \texttt{openjdk:8-jdk} as the base image.
First, POBS generates an augmented base image called \texttt{openjdk-pobs:8-jdk}. This new base image contains the POBS observability and fault injection modules. Second, the developers do a single line change in their application Dockerfile to increase observability: they replace the instruction \texttt{FROM openjdk:8-jdk} with \texttt{FROM openjdk-pobs:8-jdk}, resulting in an augmented application image. After initializing a container with this augmented application image, the fault injection module outputs a configuration file which contains the information of fault injection points. Then, the developers can configure the fault injection module via this configuration file to activate different fault injection points at runtime, so that a specific type of failure is injected into the container. By actively injecting failures and monitoring the container, developers gain the ability to evaluate how resilient the application is with respect to the injected failures. \reviseadd{If developers were to achieve a similar goal without POBS, they would have to either write extra code for observability, which would then be compiled into \texttt{my-app.jar} (line 3 in \autoref{lst:dockerfile}), or add extra libraries for monitoring and fault injection by adding more lines of instructions in the Dockerfile. Both of these two approaches are manual and require much effort in debugging and maintenance. POBS brings the benefits of adding observability discussed in \autoref{sec:observability-for-re} in an automated manner.}

\subsection{Design of POBS}

\begin{figure*}
\centering
\includegraphics[width=17.5cm]{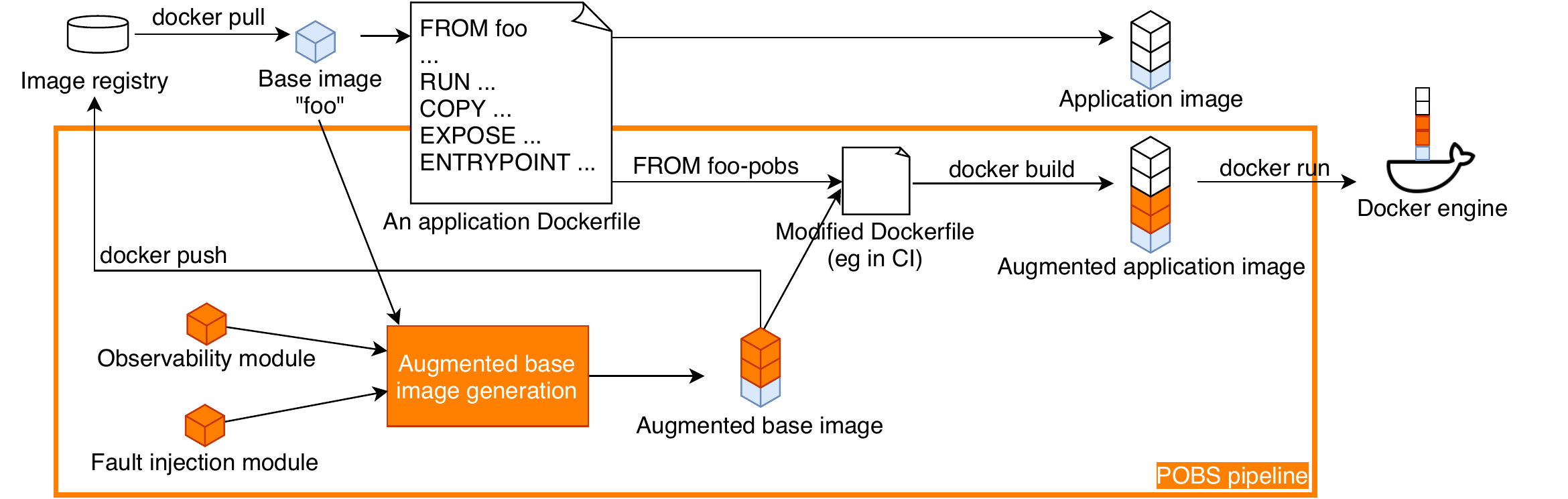}
\caption{The POBS pipeline for Docker image augmentation}\label{fig:pipeline}
\end{figure*}

The POBS pipeline extends the standard Docker workflow introduced in \autoref{sec:background-docker}. The POBS pipeline, presented in \autoref{fig:pipeline}, contains one procedure for base image augmentation (shown as an orange box in \autoref{fig:pipeline}) and one module each for observability and fault injection (shown as orange cubes in \autoref{fig:pipeline}).

The augmented base image generation procedure, discussed in \autoref{sec:base-image-generation}, adds the observability and the fault injection modules on top of a base image. The observability module improves the observability of an application container by monitoring different kinds of information such as JVM metrics, web and database transactions, etc. (more details in \autoref{sec:observability-module}). The fault injection module is able to actively inject runtime failures into the application for assessing resilience, focusing on Java exceptions \cite{zhang2019tripleagent}. It triggers the error-handling code of the application on purpose, so that the observability module can capture some abnormal behavior (more details in  \autoref{sec:fi-module}).

The key design principle for POBS is to minimize the impact on the regular workflow of the development of containerized applications. First, incorporating POBS requires minimal manual effort: the modification of one single line in the Dockerfile. Second,
POBS takes care of keeping the original dependencies needed by the application in the base image (shown as a blue cube in \autoref{fig:pipeline}). Hence, the augmented application image still meets, by construction, the functionality requirements.

Today, many developers use continuous integration and deployment (CI/CD) techniques to automatically test their applications and deliver production application images. It is possible to make POBS a part of the CI/CD pipeline, so that, for every change, an augmented application image with the same functionality but improved observability is built and deployed into production.

\subsection{Augmented Base Image Generation}\label{sec:base-image-generation}

As discussed in~\autoref{sec:observability-for-re}, conducting resilience assessment experiments requires observability and fault injection capabilities. The augmented base image generator is designed to address this issue by automatically adding observability and fault injection capabilities into Docker base images. The first step to build an application Docker image is to specify a base image in its Dockerfile (cf. ~\autoref{sec:background-docker}). Hence, POBS augments the base image declared in an application Dockerfile, in order to add observability. 


\begin{lstlisting}[caption=The generated Dockerfile for augmented image \texttt{openjdk-pobs:8-jdk}, label={lst:augmented-dockerfile}]
FROM openjdk:8-jdk
RUN apt-get install ...
COPY ./observability_module/ /home/
COPY ./fault_injection_module/ /home/
RUN mkdir /home/logs && chmod -R a+rw /home/logs
ENV FI_MODE throw_e
...
EXPOSE 4000
\end{lstlisting}

For instance, to add fault injection and observability features to the Dockerfile in \autoref{lst:dockerfile}, POBS creates an augmented base image \texttt{openjdk-pobs:8-jdk} from the original base image \texttt{openjdk:8-jdk} for the application. This augmentation is illustrated in \autoref{lst:augmented-dockerfile}.
The instructions added by the generator are used to 1) copy the files of the observability module and the fault injection module, 2) create folders to save outputs such as logs, 3) set up a series of environment variables which are necessary for setting up fault injection experiments, and 4) expose some ports so that developers are able to access a monitoring dashboard for a running container. POBS then builds an augmented image with the tag \texttt{openjdk-pobs:8-jdk} using these transformations. Developers only need to replace the base image in the original Dockerfile with the augmented one. The transformed application Dockerfile, which uses the augmented base image, is shown in \autoref{lst:transformed-dockerfile}. The sole difference between this Dockerfile and the original one in \autoref{lst:dockerfile} is highlighted in green.

\begin{lstlisting}[caption=A transformed application Dockerfile using a base image augmented by POBS, label={lst:transformed-dockerfile}]
FROM openjdk(*@\hlgreen{-pobs}@*):8-jdk
RUN install-dep.sh
COPY my-app.jar /
...
EXPOSE 8080
ENTRYPOINT ["java", "-jar", "my-app.jar"]
\end{lstlisting}

There are many popular base images for Java (cf. ~\autoref{sec:empirical-study-results}). In order to automatically generate augmented base images for each of them, our generator addresses the following two challenges.

\emph{Package manager identification} The first challenge is that these Docker images have different package managers installed. For example, the image \texttt{openjdk:8-jdk-alpine} uses \texttt{apk}~\footnote{\url{https://wiki.alpinelinux.org/wiki/Alpine_Linux_package_management}} as package manager. Others use \texttt{apt}~\footnote{\url{https://en.wikipedia.org/wiki/APT_(software)}}, typically found in the Debian/Ubuntu-based images. POBS is able to infer the correct package installation command (line 2 in \autoref{lst:augmented-dockerfile}) for different base images.

\emph{User access control} The second challenge is that, for some base images, a specific user runs the installation commands instead of the root user. This may prevent the generator from installing necessary tools and copying files. POBS addresses this challenge by checking the user name used in the base image first. If the username is not \texttt{root}, POBS adds an instruction \texttt{USER root} first when augmenting the base image. After running all the commands requiring root, the user is switched back to the original one. This ensures that the application runs with the same user name and group as intended, mitigating permission or security concerns in the augmented application image.

Technically, the POBS base image generator addresses these two challenges through a case-based strategy for Dockerfile transformation. As described in \autoref{tab:case-based-strategy}, the generator selects a sequence of instructions to be injected from a dictionary according to the original base image name and tag. For example, given a base image \texttt{foo:bar}, the generator searches for instructions for image name \texttt{foo}, tag \texttt{bar} first. If the tag is not specified in the Dockerfile, the generator searches for generic instructions for image \texttt{foo}. Finally, it uses a default instruction sequence to generate an augmented base image. By providing different instructions for different images and tags, the generator is able to use a corresponding package manager for dependency installation. If base image \texttt{foo:bar} uses a different username and group instead of root, extra instructions about switching users are added to solve permission issues.

\begin{table}
\centering
\caption{Case-based strategies in the base image generator}\label{tab:case-based-strategy}
\scriptsize
\begin{tabular}{lp{1.5cm}p{5cm}}
\toprule
Priority& Match Pattern& Usage\\
\midrule
3& ImageName:Tag& Specific snippet for one Dockerfile which uses imageName:Tag as a base image.\\
2& ImageName& Specific snippet for all Dockerfiles which use ImageName as a base image and do not match pattern 3.\\
1& Default& All the other Dockerfiles which do not match the above templates.\\
\bottomrule
\end{tabular}
\end{table}

\subsection{Observability Module}\label{sec:observability-module}

In order to evaluate the error-handling capabilities of an application, advanced monitoring is essential \cite{Dobson:UnifiedQosSLAOntologies:2006,google-monitoring-doc}. \reviseadd{Inspired by the best practices for multi-level monitoring researched by both academia \cite{Dobson:UnifiedQosSLAOntologies:2006} and industry \cite{google-monitoring-doc}, the POBS observability module provides four types of monitoring:} 1) OS layer metrics such as disk usage and system CPU load, 2) JVM runtime metrics like heap memory usage and garbage collection times, 3) library-specific metrics such as HTTP metrics and database metrics, and 4) application-specific metrics. \autoref{tab:observability-metrics} summarizes those four categories.
For example, for a web application, one library-specific metric is the number of \texttt{200} response codes per second and an application-specific metric could be the number of user logins per second. To implement the latter, the developers would specify the events which need to be instrumented, and what information needs to be collected such as method parameters and return value.

\begin{table}
\centering
\caption{The four layers of observability in POBS}\label{tab:observability-metrics}
\scriptsize
\begin{tabularx}{\columnwidth}{lX}
\toprule
Category& Metric\\
\midrule
OS Metrics& Disk usage, System CPU load, etc.\\
\midrule
JVM Runtime Metrics& Heap memory usage, Process CPU load, Loaded class count, Garbage collection count and time, etc.\\
\midrule
Library-specific Metrics& HTTP metrics (response code, response time, etc.), Database metrics (avg transaction time, transaction errors, etc.), \\
\midrule
Application-specific Metrics& Developers define them (eg. the number of video streams per second at Netflix), and the observability module automates and centralizes collection \\
\bottomrule
\end{tabularx}
\end{table}

Additionally, the observed data can be queried. The observability module provides an HTTP API to query the monitored metrics. By sending a POST request to the observability module, together with the metric name, start time and end time, the module responds with a series of values for the metrics in JSON format.

\subsection{Fault Injection Module}\label{sec:fi-module}

The fault injection module is designed to inject different kinds of faults into a Java application. This module instruments Java bytecode when a class is loaded into the JVM. The injector interface enables the developers to implement different fault models. By default, the perturbation injection module injects checked exceptions~\footnote{A checked exception in Java is an exception that must be either handled in a try-catch block, or declared to be thrown in the method where it could occur.}. 
If a method declares that it can throw exceptions, the fault injection module identifies it as an instrumentation target. As shown in \autoref{lst:perturbation-model}, method \texttt{foo()} declares that exceptions \texttt{EA} and \texttt{EB} can be thrown. Thus there are two fault injection points in this method, and each of them is controlled separately.

\begin{lstlisting}[language=Java, caption={The default perturbation model in the fault injection module},label=lst:perturbation-model]
void foo() throws EA, EB {
  // injected code when this class is loaded
  PAgent.throwExceptionPerturbation(key1);
  PAgent.throwExceptionPerturbation(key2);
  ...
}
\end{lstlisting}

For the fault injection module to work properly, some options need to be set up when the application is started up. The fault injection module takes options by extracting different environment variables. In this way, developers are able to configure the fault injection module by giving extra environment variables such as \texttt{docker run -e foo=bar ...}. The complete list of supported configurations is presented in \autoref{tab:fi-options}. Option \texttt{FILTER} and \texttt{EFILTER} control the range of bytecode instrumentations done by the fault injection module. By defining these two options, fault injection experiments could be focused on specific packages and exception types. Option \texttt{MODE} is used to specify which perturbation model will be used by the fault injection module. If developers implement their own perturbation mode, they could activate it by changing this option. Option \texttt{INJECTPOSITION} defines the position where to inject an exception in a method. By default, the module injects exceptions at the beginning of a method, which is an extreme case for the method: the whole method body is short-circuited. Option \texttt{RATE}, \texttt{DEFAULTMODE} and \texttt{COUNTDOWN} are set to control the behavior of each perturbation injector.

The fault injection module, together with the observability module, enables developers to conduct fault injection experiments for resilience assessment. When a fault is injected by the fault injection module, it may or not have an influence on the metrics being monitored by the observability module. To detect whether a specific injected fault leads to the application's abnormal behavior, POBS uses causal impact analysis. The principle of causal impact analysis is: 1) it considers a period of monitoring data without any fault injection as input, 2) it analyzes the monitoring data and formalizes a prediction model, 3) it compares the values of the monitored metric during fault injection experiments against the predicted values, and 4) if a significant difference between the real values and the prediction is detected, the injected failure is considered to have a causal impact on the monitored metric.

\begin{table}
\centering
\caption{Options for the fault injection module}\label{tab:fi-options}
\scriptsize
\begin{tabular}{lp{5.7cm}}
\toprule
Configuration Variable& Default Value and Usage\\
\midrule
FILTER& ".*", package names to be instrumented with bytecode transformation.\\
EFILTER& ".*", methods which declare to throw the matched exception types are instrumented.\\
RATE& "1", the probability of exception injections.\\
MODE& "throw\_e", the mode to be used for fault injection experiments.\\
INJECTPOSITION& "0", inject exceptions after a number of instructions in a method. "0" means the beginning of a method.\\
DEFAULTMODE& "off", whether to inject exceptions as soon as the fault injection point is reached.\\
CSVPATH& "logs/perturbationPointsList.csv", path to a file which stores the detected fault injection points.\\
COUNTDOWN& "1", the maximum number of thrown exceptions for each fault injection point.\\
\bottomrule
\end{tabular}
\end{table}

\subsection{Implementation}

The POBS base image generator and all the orchestration scripts are written in Python 3. The fault injection module is an enhanced version of TripleAgent \cite{zhang2019tripleagent}, which allows configuring fault injection via environment variables. \reviseadd{As introduced in \autoref{sec:fi-module}, TripleAgent works as a Java agent that is attached to the JVM when an application is started. It uses the ASM library\footnote{\url{http://asm.ow2.org}} to instrument Java bytecode with fault injection points for methods which declare to throw exceptions.} The observability module is implemented using Glowroot~\footnote{\url{https://glowroot.org/}}, a low-overhead application performance management tool for JVM. For causal impact analysis, we use Google's library CausalImpact \cite{causalimpact} implemented in Python. For the sake of open-science, the code is made publicly available at \url{https://github.com/KTH/royal-chaos/tree/master/pobs}.

\section{Experimental Methodology}
\label{sec:evaluation}

The evaluation of POBS is based on the following research questions.

\newcommand\rqfunctionality{To what extent do base images, augmented by POBS, preserve the functionality of the original Docker base images?}

\newcommand\rqobservability{To what extent are base images augmented by POBS successful at increasing observability of real-world Dockerized Java applications?}

\newcommand\rqfaultinjection{To what extent are augmented base images useful for fault injection analysis of real-world Dockerized Java applications?}

\newcommand\rqoverhead{What is the impact on  performance of the augmented base image?}

\begin{itemize}
    \item RQ1: \rqfunctionality
    \item RQ2: \rqobservability
    \item RQ3: \rqfaultinjection
    \item RQ4: \rqoverhead
\end{itemize}

\subsection{Methodology for RQ1}\label{sec:methodology-rq1}
\label{sec:functionality-protocol}

In order to test whether the observability module and the fault injection module are successfully integrated into a base image, we perform two checks:
1) whether a `docker build' of the augmented image succeeds
2) whether a simple Java application can run successfully on top of this augmented base image, in two modes. 

For the second check, we need an application. For RQ1, we use a demo application, called `DLIEEE` for short. This application consists of downloading the official IEEE LaTeX package on the internet. It contains one class which uses Apache's \texttt{Commons IO}~\footnote{\url{https://commons.apache.org/proper/commons-io/}} library to download the file. There is one method \texttt{downloadTheFile()} which declares that it can throw \texttt{InterruptedExceptions} and \texttt{IOExceptions}. These two exceptions are the target for the fault injection module.

We perform these two checks for all the base images in our dataset, as presented in \autoref{sec:empirical-study}. There is one DLIEEE image per base image, which is tested with the following two experiments:

\emph{Mode A} Both the observability module and the fault injection module print a message in a log to declare their successful attachment. This test passes if the file is successfully downloaded with the correct checksum.

\emph{Mode B} Both the observability module and the fault injection module are successfully attached as well. In addition, the fault injection is activated and an \texttt{IOException} is injected at the beginning of method \texttt{downloadTheFile()}. The container thus fails to download the file. This test passes if no file is downloaded and if an exception injection message is detected in the container's log output.

To sum up, an augmented base image is considered valid if and only if:
1) \texttt{docker build} succeeds;
2) DLIEEE passes the mode A test;
3) DLIEEE passes the mode B test.

\subsection{Methodology for RQ2}\label{sec:methodology-rq2}

Now we consider the application images. Since RQ2 focuses on observability improvement, the fault injection module is set to off. Next for each Dockerfile, we perform three steps.

Step 1, Generate an augmented base image for the Dockerfile:
The POBS base image generator, presented in \autoref{sec:base-image-generation}, builds an augmented version for the base image of the Dockerfile under study. The new augmented base image contains the observability module and the fault injection module, which are used in the following steps.

Step 2, Build the Dockerfile using the augmented base image:
If an augmented base image is successfully created, the original application Dockerfile is updated by replacing the \texttt{FROM} instruction, e.g. `FROM openjdk` to `FROM openjdk-pobs`. Then the \texttt{docker build} command is used to build the augmented application image.

Step 3, Run the container:
In the final step of the experiment we launch \texttt{docker run} to execute the Java application in a container and we check that the observability module is working correctly. We consider that observability is improved if the following criteria hold: 1) the observability module outputs a successful attachment message in the container log, 2) the operating system metrics, JVM runtime metrics, and library-specific metrics, which are presented in \autoref{tab:observability-metrics}, can be extracted via the observability module API, and 3) \reviseadd{the container does not crash after its initialization. As these containers are designed to be continuously running, the maximum duration of each experiment is defined as $1$ minute. If all of the three criteria above hold for $1$ minute, step 3 is considered successful and the next experiment starts.}

\subsection{Methodology for RQ3}\label{sec:methodology-rq3}
In RQ3, we study whether augmented base images are useful for fault injection analysis.
To answer RQ3, we consolidate application-specific knowledge for deploying an application, generating specific workload, and checking the correctness of the application's behavior. We carefully select three applications from our dataset according to the following selection criteria: 1) each application contains at least one Dockerfile which passes all the observability tests defined in \autoref{sec:methodology-rq2}, because fault injection analysis requires observability, 2) size and popularity per \autoref{fig:covered-projects} are above average, and 3) the general usage of the application is sufficiently documented in the GitHub repository.
According to the criteria above, we select Grobid\footnote{\url{https://github.com/kermitt2/grobid}}, Eclipse hawkBit\footnote{\url{https://github.com/eclipse/hawkbit}}, and Spotify Heroic\footnote{\url{https://github.com/spotify/heroic}} as the experimental targets for fault injection. Grobid is a tool that extracts information from scholarly documents. Eclipse hawkBit is a software distribution management back-end for IoT devices. Spotify Heroic is a scalable time series database. Descriptive information for all selected projects is presented in \autoref{tab:projects-for-rq3}.

\begin{table}
\centering
\caption{Descriptive Statistics of The Applications for RQ3}\label{tab:projects-for-rq3}
\scriptsize
\begin{tabularx}{\columnwidth}{Xrrrrr}
\toprule
Application& Version& LoC& Stars& Commits& Contributors\\
\midrule
Grobid& 0.6.0& 690K& 1.2K& 1.9K& 38\\
HawkBit& 0.3.0M6& 126K& 275& 2.3K& 50\\
Heroic& 2.1.0& 67K& 834& 1.3K& 34\\
\bottomrule
\end{tabularx}
\end{table}

The fault injection experiments are conducted according to the following steps. First, we try to build the application in order to get the necessary application jar files (same step as in \autoref{sec:evaluation-dataset}). Second, the POBS base image generator is applied to build an augmented base image, which is used to build an augmented application image. Then by initializing an application container and feeding it with the workload, we get all the fault injection points information from the fault injection module. 

\reviseadd{The workloads are constructed to cover a frequently-used feature in each application, according to its documentation.}
For Grobid, the workload consists of passing the PDF of a research paper to Grobid's \texttt{processHeaderDocument} API. This API returns the main metadata of the input PDF document in an XML format \cite{GROBIDDocumentation}. 
The workload for HawkBit contains two RESTful API invocations: one is a POST request that adds the basic information about a software module to be deployed, and the other one is a GET request that retrieves all the registered software modules. 
The workload for Heroic contains two requests as well: two data points are first inserted into the database, and another request then tries to query the inserted points.

For each fault injection point, we first keep the fault injection point deactivated and exercise the container with a workload for $5$ minutes. Then a specific fault injection point is activated, and the same workload is used as input for another $5$ minutes. Then, the 'normal' run is used as a reference to identify abnormal behavior.

There are two different kinds of abnormal behavior to be analyzed by the experiments: 
1) correctness: every response from the API is compared to a reference response, to calculate whether the API still behaves properly under fault injection;
2) performance: a causal impact analysis is conducted on the monitored performance metrics, to detect whether the metrics are impacted by the injected failures.

By querying the monitoring module via its HTTP API, the usage of CPU resources in JVM is exported as an example for causal impact analysis of performance issues. The monitoring data is saved in a JSON file which contains a set of timestamp-value pairs. The timestamp at which the fault injection module begins to inject exceptions is also provided for impact analysis. Then the monitoring data is used as input to Google's CausalImpact tool. CausalImpact outputs two values as an analysis result: 1) the probability of obtaining a causal impact by chance in the posterior area, denoted as \texttt{p-value}, and 2) the relative effect on average in the posterior area, denoted as \texttt{re}. \texttt{re} is a signed percentage that describes how the metric changes in the posterior area. For example, $50\%$ means the monitoring metric is increased by $50\%$ on average in the posterior area. A smaller \texttt{p-value} from CausalImpact indicates that \texttt{re} is more statistically significant. A larger \texttt{re} signifies a more severe impact from the fault injection point. \reviseadd{Note that performing many rounds of experiments could potentially increase the risk of false positives (due to data dredging). Thus the fault injection experiment of each point is  executed once.}

\subsection{Methodology for RQ4}
We evaluate the performance overhead caused by base image augmentation, by running the same experiment targets and workload as RQ3. The overhead is evaluated at three different levels: the image level, the running container level, and the application level. At the image level, we measure the disk size increase of the image. This is queried by the command \texttt{docker images}. At the container level, we capture the CPU usage and memory usage. 
At the application level, we measure the average response time of each application's APIs. Two runs are made: a normal execution using the original application image (reference group), and an execution using the augmented base image without injecting any failure. We do not evaluate the performance overhead during fault injection because the abnormal behavior caused by injected failures may be related to performance. All executions are made for 5 minutes, including $300$ calls to each API. For statistical purposes, the same measurement is collected 30 times to calculate the average \cite{Arcuri:A_practical_guide_for_using_statistical_tests_to_assess_randomized_algorithms_in_software_engineering}.


\section{Experimental Results}

We now present our experimental results based on the methodologies presented in \autoref{sec:evaluation}. RQ1 focuses on Docker base images, RQ2 and RQ4 on application images, and RQ3 on a fault injection case study.

\subsection{RQ1. Creating POBS base images}

The dataset introduced in \autoref{sec:evaluation-dataset} includes \nbApplicationDockerFiles  Dockerfiles located in \nbProjects projects. This represents  \nbBaseImages unique base images. The POBS pipeline can create a valid augmented base image for $72$ of these \nbBaseImages base images.

The augmented base images passed the evaluation tests mentioned in \autoref{sec:functionality-protocol}. They successfully maintain the original functionalities and provide new monitoring and fault injection capabilities. In order to contribute useful, novel artifacts to the research and Docker community, we have publicly released, on Docker Hub, the augmented based images for the $25$ most popular base images of our dataset. Overall, this first experiment shows the wide applicability of our approach.

Let us consider the most popular base image, \texttt{java:8}, as an example here. The application ``DLIEEE'' uses \autoref{lst:dlieee-dockerfile} to build its Docker image. POBS pipeline successfully builds an augmented base image called \texttt{java-pobs:8} first. Then the augmented base image is used to build and run a ``DLIEEE'' container. This container passes test experiments with both mode A and mode B described in \autoref{sec:methodology-rq1}, which means both the observability module and the fault injection module are successfully attached and the fault injection module succeeds in injecting the specific exceptions. Compared to the original ``DLIEEE'' application image, the augmented image provides both the extra observability and the capability of conducting fault injection experiments.

\begin{lstlisting}[caption={The DLIEEE Dockerfile using \texttt{java:8} as the base image},label=lst:dlieee-dockerfile]
FROM java:8
WORKDIR /root
# install openjdk8 if Java is not installed
COPY ./install_openjdk8.sh /root
RUN ./install_openjdk8.sh
# run the application
COPY ./DLIEEE.jar /root
ENTRYPOINT ["java", "-jar", "DLIEEE.jar"]
\end{lstlisting}

We now analyze the $14$ failure cases, where our approach did not yield valid POBS base images. The most common failure, which occurs in $9$ out of $14$ failure cases, is related to the Java version used in the base image. Since the POBS fault injection module leverages some capabilities that are available only in Java 8+, we cannot support base images with Java 7 or older.

The second most frequent cause for a failure ($3$ out of $14$) is an invalid reference format error. This error means that in the original Dockerfile, variables are used to describe the base image name and tag. For example, instruction \texttt{FROM jenkins/jenkins:\$jenkins\_tag} needs developers to specify the exact tag to be used. When POBS automatically augments base images, these environment variables are left as blank, which causes the reference format error. This could be resolved by specifying the correct environment variables, or make a default tag name as \texttt{latest}.

Finally, there are $2$ failed cases which are caused due to denied permissions. These two base images have different permissions set up for creating folders and changing the owner of folders. This can be resolved by adding more templates to specifically handle such base images.

\begin{mdframed}[style=mpdframe,nobreak=true,frametitle=Answer to RQ1]
The base image generator of POBS is effective: it successfully generates augmented base images for $72/86$ (84\%) base images collected from the field. This is arguably a high coverage, which could be further improved with more engineering (e.g., by adding new templates for less common base images). \reviseadd{These automatically augmented base images enable developers to enhance the observability of an application with minimal effort.}
\end{mdframed}

\subsection{RQ2. Improving Observability in Application Images}

We now discuss the extent to which our technique can add observability capabilities into Java application images.
In total, there are \nbApplicationDockerFiles application Dockerfiles under evaluation in this research question (all those Dockerfiles that can be built using a default \texttt{docker build} command, see \autoref{sec:evaluation-dataset}). We could run a correct build for up to $87\%$ of these Dockerfiles ($148/\nbApplicationDockerFiles$) with augmented base images.

We observe one case where an augmented base image is successfully built and passes the tests in RQ1 while the application fails to build an image on top of this augmented base image. The reason for this failure is Maven's \texttt{OutOfMemoryError}, which means the JVM running Maven has run out of memory since now there are more components attached to the Java process.

Next, we execute a \texttt{docker run} command on each of the $148$ application images that include a POBS augmented base image. This command is successful in $131$ cases ($77\%$ of \nbApplicationDockerFiles Dockerfiles), i.e., the application image: 
1) is successfully attached to the observability module, 
2) can continuously run for one minute without exiting exceptionally, and 
3) is able to give generic monitoring metrics via the monitoring API. 

There are $17$ augmented application images that fail the verification criteria mentioned above. The most common failure, which occurs in $10$ cases, is that the JVM running in these images fails to recognize the necessary environment variable which declares the POBS components. The remaining $7$ failed cases are caused by permission errors. For example, the POBS observability module fails to create necessary files in the target container.

Recall that RQ1 focuses on the number of unique base images while RQ2 focuses on the number of application Dockerfiles. \autoref{tab:summary-of-rq1-and-rq2} summarizes the relation between these two different aspects. It shows that no matter which dimension is chosen, the percentage of successful cases is comparable. The idea of automatically augmenting Docker base images works for an arguably high percentage of Dockerized Java applications.

\begin{table}
\centering
\caption{The relation between unique base images and application Dockerfiles in RQ1 and RQ2}\label{tab:summary-of-rq1-and-rq2}
\scriptsize
\begin{tabularx}{\columnwidth}{lRRR}
\toprule
Dimension& Base Image Augmentation& Build Success& Run Success\\
\midrule
\nbBaseImages unique base images& 72 (83.7\%)& 71 (82.6\%)& 56 (65.1\%)\\
\nbApplicationDockerFiles Dockerfiles&149 (87.6\%)&148 (87.1\%)&131 (77.1\%)\\
\bottomrule
\end{tabularx}
\end{table}

\begin{mdframed}[style=mpdframe,nobreak=true,frametitle=Answer to RQ2]
There are $148/\nbApplicationDockerFiles$ ($87\%$) Docker application images that could be constructed with augmented Docker base images. With a default execution command, we can prove that $131/\nbApplicationDockerFiles$ ($77\%$) augmented application images are runnable. This shows that our approach improves observability in a fully automated manner. Developers have virtually no change to make their Dockerized Java applications benefit from automated monitoring. \reviseadd{Automatically improving the observability at the image level can also be done in continuous integration, since the only task to be performed is updating the base image declarations.}
\end{mdframed}

\subsection{RQ3. Results on Fault Injection}

Out of the considered three applications, the fault injection module identifies $75$ fault injection points, all covered by the experiment workload.
By conducting one fault injection experiment per covered fault injection point, a resilience analysis report is computed. We summarize it in \autoref{tab:results-summary-rq3}. Each row in the table summarizes the results for one application, including the number of total potential fault injection points, the covered ones, the verified resilient points, and the number of points that cause a visible performance impact. A resilient point means that during the fault injection experiment, the application still gives correct responses according to the given workload. A performance issue means that the causal impact analysis module reports the usage of CPU resources is statistically different when the fault injection point is activated (p-value < 0.05).

\begin{table}[!tb]
\centering
\scriptsize
\caption{The Summary of Experimental Results on Grobid, HawkBit and Heroic}\label{tab:results-summary-rq3}
\begin{tabularx}{\columnwidth}{lrrrr}
\toprule
\textbf{App.}& \textbf{Total FI Points}& \textbf{Covered}& \textbf{Resilient}& \textbf{Performance Issues}\\
\midrule
Grobid& 100& 16& 6& 5\\
HawkBit& 73& 10& 0& 0\\
Heroic& 214& 49& 4& 8\\
\midrule
TOTAL& 387& 75& 10& 13\\
\bottomrule
\end{tabularx}
\end{table}

\subsubsection{Experiment Results on Grobid}
There are $16$ points out of $100$ in total that are covered by the workload defined in \autoref{sec:methodology-rq3}. By conducting fault injection experiments on each covered point, it shows that there are $6$ points that are resilient to the injected exceptions. The experiments also find that $5$ fault injection points have an obvious impact on the application's performance (e.g., the usage of CPU resources in this experiment). The results of Grobid are presented in \autoref{tab:results-grobid-rq3}. Every row records the basic information of a fault injection point including its class name, method name, and thrown exception type. A row also presents the correctness rate of all the responses, the probability of obtaining a performance causal impact by chance in the posterior area \texttt{p-value}, together with the relative effect on average in the posterior area \texttt{re}.

\begin{table*}
\centering
\scriptsize
\caption{Resilience Evaluation for Docker Allowed by POBS (Column P-value stands for the probability of obtaining an effect by chance. Column RE stands for the relative effect on average in the posterior area.)}\label{tab:results-grobid-rq3}
\begin{tabularx}{\textwidth}{Xlllrrr}
\toprule
No.& Full Class Name& Method Name& Exception Type& C. Rate& P-value& RE\\
\midrule
1& o/g/core/factory/GrobidPoolingFactory& destroyObject& java/lang/Exception& 100\%& 0.35& 5.21\%\\
2& o/g/core/factory/GrobidPoolingFactory& passivateObject& java/lang/Exception& 100\%& 0.31& -7.82\%\\
3& o/g/core/sax/PDFALTOAnnotationSaxHandler& endElement& org/xml/sax/SAXException& 100\%& 0.38& 2.30\%\\
4& o/g/core/sax/PDFALTOAnnotationSaxHandler& startElement& org/xml/sax/SAXException& 100\%& 0.38& 3.29\%\\
5& o/g/core/sax/PDFMetadataSaxHandler& endElement& org/xml/sax/SAXException& 100\%& 0.44& 2.23\%\\
6& o/g/core/sax/PDFMetadataSaxHandler& startElement& org/xml/sax/SAXException& 100\%& 0.31& -7.25\%\\
7& o/g/core/factory/GrobidPoolingFactory& activateObject& java/lang/Exception& 0\%& <0.01& -62.39\%\\
8& o/g/core/factory/GrobidPoolingFactory& makeObject& java/lang/Exception& 0\%& <0.01& -47.30\%\\
9& o/g/core/engines/AuthorParser& processingCitation& java/lang/Exception& 0\%& 0.11& 15.21\%\\
10& o/g/core/engines/HeaderParser& processingHeaderBlock& java/lang/Exception& 0\%& 0.01& -30.92\%\\
11& o/g/core/sax/PDFALTOSaxHandler& endElement& org/xml/sax/SAXException& 0\%& <0.01& -49.26\%\\
12& o/g/core/sax/PDFALTOSaxHandler& startElement& org/xml/sax/SAXException& 0\%& 0.02& -34.09\%\\
13& o/g/core/features/FeaturesVectorName& addFeaturesName& java/lang/Exception& 0\%& 0.48& 1.12\%\\
14& o/g/core/features/FeaturesVectorAffiliationAddress& addFeaturesAffiliationAddress& java/lang/Exception& 0\%& 0.29& 8.48\%\\
15& o/g/core/features/FeaturesVectorCitation& addFeaturesCitation& java/lang/Exception& 0\%& 0.15& 15.70\%\\
16& o/g/core/features/FeaturesVectorDate& addFeaturesDate& java/lang/Exception& 0\%& 0.33& 4.37\%\\
\bottomrule
\end{tabularx}
\end{table*}

Let us take the first fault injection point (No. 1) in \autoref{tab:results-grobid-rq3} as an example. The fault injection point is located in class \texttt{GrobidPoolingFactory}, at the beginning of method \texttt{destroyObject}. When the fault injection point is activated, it keeps injecting \texttt{Exception} every time the point is reached. However, even injecting these exceptions when a PDF is sent to Grobid, 100\% of the responses are correct according to the reference output. The p-value of observing a performance impact when activating the fault injection point is $0.35$. The relative impact, on average, of this fault injection point is $5.21\%$. In this case, it means that CausalImpact does not have enough statistical evidence to claim that the injected exceptions have a causal impact on the CPU usage. This is good news for the developers of Grobid, because the injected failures do not prevent Grobid from correctly extracting the header information from a PDF file. There is no obvious performance overhead detected during the fault injection either.

\begin{figure}
    \centering
    \subfloat[Normal Execution\label{fig:grobid-causal-impact-reference}]{\includegraphics[width=8.5cm, clip=true, trim=30mm 60mm 30mm 60mm]{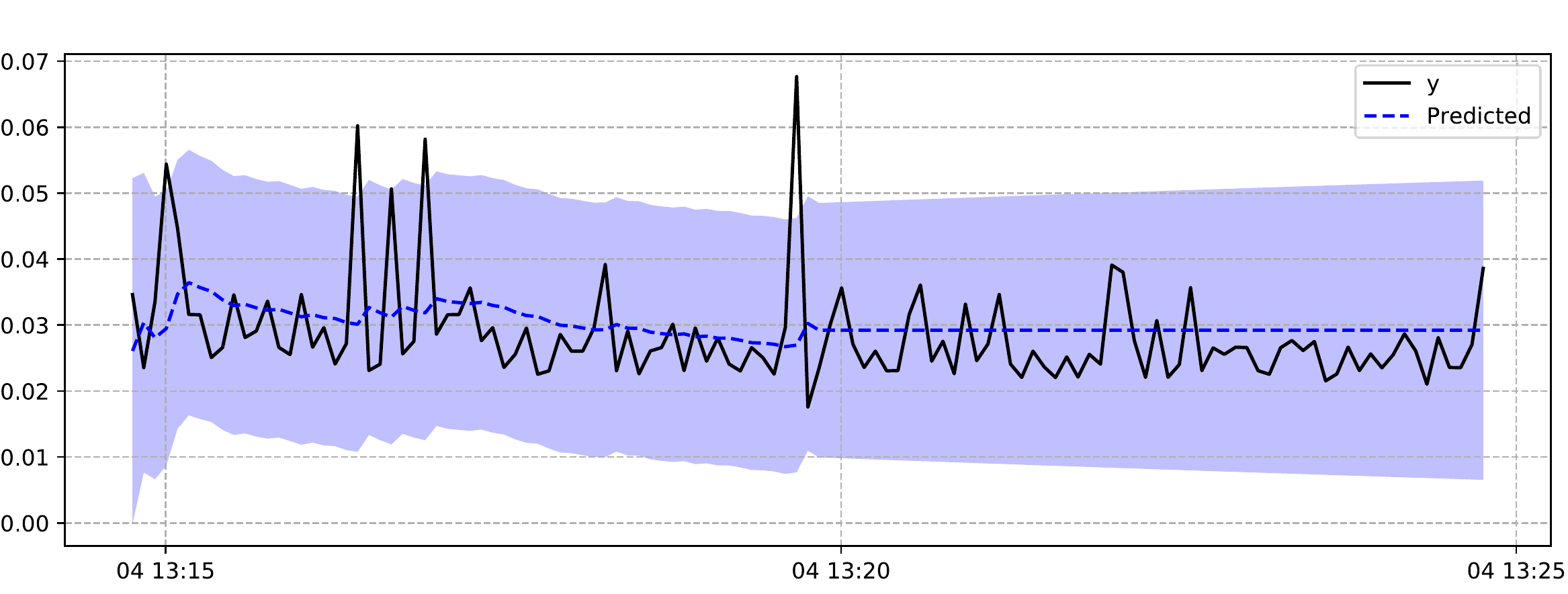}}
    \\
    \subfloat[Fault Injection for Dockerized Java: the vertical dashed line shows when the exceptions starts to be injected.\label{fig:grobid-causal-impact}]{\includegraphics[width=8.5cm, clip=true, trim=30mm 60mm 30mm 60mm]{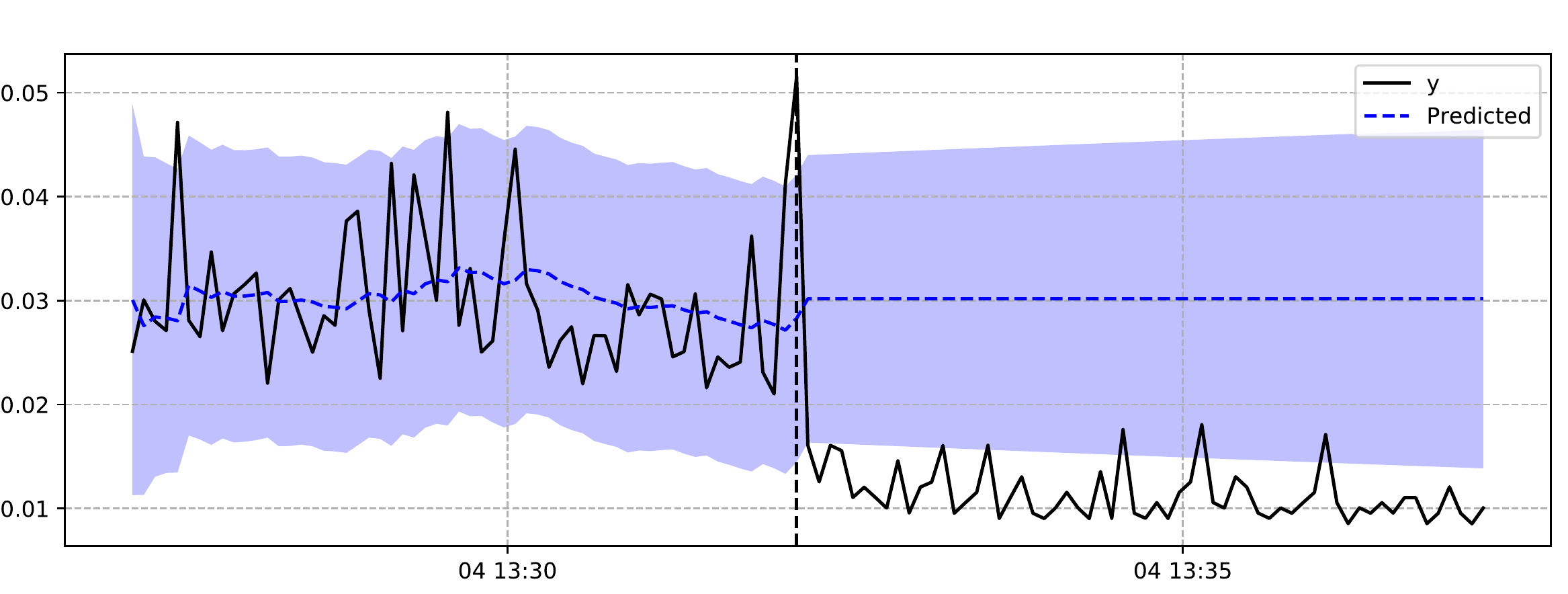}}
    \caption{Causal impact analysis of the CPU usage for point no. 7 in Grobid. The application is monitored for 10 minutes. The black solid line is the real value of the heap memory usage. The blue dashed line shows the predicted trend of the metric.}
\end{figure}

To support engineers in understanding a causal impact on performance metrics, CausalImpact generates a visualization of the monitored information. \autoref{fig:grobid-causal-impact-reference} and \autoref{fig:grobid-causal-impact} show the visualization of the application thread's CPU usage data for the normal execution and fault injection experiments of fault injection point No. 7 in \autoref{tab:results-grobid-rq3}. Axis X records wall-clock time. Axis Y shows the percentage value of the CPU usage. The black solid line is the real value of the monitored data. The blue dashed line shows the predicted trend of the metric. The visualization of the normal execution acts as a reference group. \autoref{fig:grobid-causal-impact} shows that the container is injected with exceptions after the vertical dashed line. There is a clear relation between the prediction and the real monitored data. Thus an \texttt{Exception} which is injected at the beginning of method \texttt{activateObject}, class \texttt{GrobidPoolingFactory} does cause a causal impact on the usage of CPU resources: the thread's CPU usage significantly drops during the fault injection experiment. According to \autoref{tab:results-grobid-rq3}, the effect is also quantified by the causal impact analysis. The  p-value of this point less than $0.01$, with a relatively higher impact ($-62.39\%$ on average) on the usage of CPU. The CPU usage significantly drops when method \texttt{activateObject} is perturbed with exceptions, as other methods called from \texttt{activateObject} are not called anymore.

\subsubsection{Experiment Results on HawkBit}
As summarized in \autoref{tab:results-summary-rq3}, HawkBit has $73$ potential fault injection points in total. From them, $10$ injection points are covered by the workload. Out of these $10$ points, $2$ points are located in package \texttt{org/eclipse/hawkbit/security}. Another $4$ points are in \texttt{org/eclipse/hawkbit/rest}. The rest of the $4$ points are in \texttt{org/eclipse/hawkbit/autoconfigure}. These points are in methods \texttt{doFilterInternal} or \texttt{doFilter}. After manually checking the source code of these methods, we find that all of them are related to request filtering. They declare to throw either a \texttt{ServletException} or an \texttt{IOException}. The fault injection experiment allowed by POBS is a success: it verifies that exceptions in the filter should result in the server providing the appropriate error code. By conducting chaos engineering experiments on these methods, developers build confidence in their behavior under error: a request that triggers an exception in the HTTP request filters cannot be responded to correctly. By conducting causal impact analysis on these fault injection points, developers also confirm that the error handling logic does not bring extra performance overhead either.

\subsubsection{Experiment Results on Heroic}
Heroic has the most covered fault injection points among the three applications: $49$ fault injection points are covered by the workload. As shown in \autoref{tab:results-summary-rq3}, by injecting the corresponding type of exceptions in each covered fault injection point, POBS identifies $4$ out of $49$ points which are resilient to the injected exceptions. There are also $8$ points identified as having a statistically significant impact on CPU usage during the fault injection experiments. Both information are useful and could not be extracted without POBS in a vanilla setup.

For example, the method \texttt{resolved} in class \texttt{ResultCollector} is marked as a resilient point with respect to a Java \texttt{Exception}. The method declares to potentially throw an \texttt{Exception}, and is therefore identified as a fault injection point by POBS. During the experiment, the fault injection module keeps injecting exceptions when the method \texttt{resolved} is invoked. By analyzing the content of the response, POBS still successfully gets the query result of the expected data points. At the same time, the probability of obtaining a causal impact by chance during this chaos engineering experiment (\texttt{p-value}) is 0.31, which is a relatively large value. A large \texttt{p-value} means that the causal impact on the usage of CPU resources during the fault injection experiment does not have a statistical meaning. In other words, the developers have designed a proper backup plan for Heroic when such exceptions happen in production. As one of the goals of chaos engineering, the application's resilience to a specific injected exception is verified by this experiment.

\begin{mdframed}[style=mpdframe,nobreak=true,frametitle=Answer to RQ3]
Our technique is able to successfully add fault injection capabilities in Dockerized Java applications, with essentially no developer effort. The combination of improved observability and fault injection enables developers to understand error handling and identify resilience issues with respect to correctness and performance. \reviseadd{With the help of POBS, developers gain more insight into the resilience of a Dockerized Java application.}
\end{mdframed}

\subsection{RQ4. Performance Overhead Evaluation}

\autoref{tab:overhead} presents the key performance metrics related to POBS. There are 4 metrics in consideration: the size of the augmented Docker image, the average CPU usage during an experiment, the average memory usage, and the average response time of the invoked API in the workload. Both the absolute values and percentages are presented in the table for comparison. According to \autoref{tab:overhead}, the augmented images which contain POBS components require 44MB extra storage, basically the size of POBS' infrastructure. The augmented application image causes an increase in CPU usage between 0.34\% and 1.57\%. The memory usage is increased by 466MB at most. The response time of an application's API is increased by at most $0.01$ seconds. Overall, this can be considered acceptable in most realistic scenarios.

Let us take Grobid as an example. The size of the original Grobid application image is $1.57$ GB. The size of the augmented application image is $1.61$ GB, which is $2.5\%$ larger than the original one. When the application is deployed with the original base image, the average CPU and memory usage are respectively $3.34\%$ and $3.84$ GB. The average response time is $0.099$ seconds. When the application is deployed with the augmented base image and the fault injection is off, the application takes $4.91\%$ CPU usage and $4.03$ GB memory usage on average, which are $47.0\%$ and $4.9\%$ larger than the reference group respectively. The average response time is $0.109$ seconds, which is $10\%$ larger than the reference group.
While increases in disk, memory, and CPU mainly impact infrastructure budgeting, the end-user mostly cares about response time: in that case the POBS overhead is virtually invisible for the end-user (within milliseconds on the considered benchmark).

\begin{table}
\centering
\caption{The performance overhead caused by POBS}\label{tab:overhead}
\scriptsize
\begin{tabularx}{\columnwidth}{llRRr}
\toprule
Application& Category& Original Image& Augmented Image& Increase\\
\midrule
\multirow{4}{*}{Grobid}& Image Size& 1569MB& 1614MB& 45MB (2.9\%)\\
& CPU Usage& 3.34\%& 4.91\%& 1.57\%\\
& Memory Usage& 3838MB& 4098MB& 200MB (4.9\%)\\
& Response Time& 0.099s& 0.109s& 0.01s (10\%)\\
\midrule
\multirow{4}{*}{HawkBit}& Image Size& 157MB& 201MB& 44MB (28\%)\\
& CPU Usage& 1.50\%& 1.84\%& 0.34\%\\
& Memory Usage& 1184MB& 1410MB& 226MB (19.1\%)\\
& Response Time& 0.0118s& 0.0122s& 0.0004s (3.4\%)\\
\midrule
\multirow{4}{*}{Heroic}& Image Size& 767MB& 811MB& 44MB (5.7\%)\\
& CPU Usage& 0.57\%& 1.06\%& 0.49\%\\
& Memory Usage& 305MB& 771MB& 466MB (153\%)\\
& Response Time& 0.0088s& 0.0090s& 0.0002s (2.3\%)\\
\bottomrule
\end{tabularx}
\end{table}

\begin{mdframed}[style=mpdframe,nobreak=true,frametitle=Answer to RQ4]
The size of the augmented Java application only slightly increases (44MB), and it does not have significant extra storage costs. The CPU usage and memory usage are higher when running the application with the augmented base image, because of the added monitoring and observability. From the user's perspective, the response time is at most $0.01$ seconds slower, which can be considered acceptable in most use cases. \reviseadd{The limited overhead indicates that POBS is promising when applied in a staging or production environment, so that developers are able to improve the resilience of their application even after its deployment.}
\end{mdframed}

\section{Discussion}

\subsection{Threats to Validity}

\emph{Internal Validity}
The main threat to internal validity is that the fault injection module only injects exceptions at the beginning of methods which declare that they can throw exceptions. The behavior of an application under fault injection may change if an exception is injected in different locations of the target method. Further work is needed to systematically analyze the fault injection searching space.

\emph{External Validity}
Per the methodology for RQ3, POBS is only evaluated with two real-world Java applications, which is a threat to the external validity. It would be interesting to select more projects in different domains and to evaluate how POBS works on these projects. The workload of our experiments could be diversified in order to trigger more fault injection points as well.

\subsection{POBS and Security}
A POBS-augmented application image contains extra components, namely the observability module and fault injection module. They can impact the security of an application. As described in \autoref{sec:base-image-generation}, POBS uses the same user name and group to run the application as the original image does. Thus the actual user that executes the application has the same permissions. Furthermore, security issues can be detected by running vulnerability scanning tools for Docker on the augmented images. To this end, we deploy a static analysis tool called Clair\footnote{\url{https://github.com/quay/clair}} to scan the augmented images of Grobid, HawkBit, and Heroic for security vulnerabilities. The results show that the augmented images have the same number of detected vulnerabilities as the original images. This suggests that POBS does not bring extra security issues to a certain extent.


\section{Related work}
Our work relates to 
the fields of containerization, observability, and fault injection.

\subsection{Containerization \& Docker}

Cito et al. \cite{7962382,Schermann:2018:SIS:3196398.3196456} conducted an extensive exploratory study of Dockerfiles on GitHub to assess their distribution across projects, conformance to quality guidelines, and evolution between revisions. Their study, like ours, consists of the analysis of Dockerfiles to find the most popular base images, as well as determining if the Dockerfiles can be successfully built. However, unlike our study, the intention of their analysis was to investigate the usage of Docker as a containerization tool and not as a potential target for improved observability and fault injection. Henkel et al. \cite{Henkel:DevOpsArtifactsForDocker} analyzed a large Dockerfile corpus \cite{henkel2020dataset} to implement an automated parsing, rule-mining, and rule-enforcement engine for Dockerfiles. Their tool is meant to help practitioners in achieving high-quality DevOps artifacts. Hassan et al. \cite{Hassan:ASE2018:RUDSEA} proposed RUDSEA, an approach that recommends updates of Dockerfiles based on analyzing changes on software environment assumptions. Oumaziz et al. \cite{Oumaziz:ICSME:8919205} summarized the practices of expert Dockerfile maintainers in order to mitigate duplicates in Dockerfiles and how to handle them. Instead of improving applications' observability like POBS, these works focus on improving the quality of Dockerfiles. Another empirical study explores official as well as community Docker images available on Docker Hub \cite{ibrahim2020too}, where Ibrahim et al. show that the diversity of Docker images offer great, and often undocumented, alternatives in terms of installed software, resource efficiency, and security vulnerabilities.

Studies have also been conducted to address issues concerning the security of using Docker. Gao et al. \cite{8023126} investigated information leaks between multi-tenant hosts and the containers running on them, which would allow system-wide information to be available to a malicious container. 
Lin et al. \cite{Lin:2018:MSL:3274694.3274720} evaluated real-world exploits, privilege escalation attacks in particular, that compromise container isolation.
Zerouali et al. \cite{DBLP:journals/corr/abs-1811-12874,Zerouali:SANER:8667984} conducted a study on the ``technical lag'' with respect to outdatedness and security vulnerabilities present in official and community Docker images based on the Debian Linux distribution.
Shu et al. \cite{Shu:2017:SSV:3029806.3029832} implemented a framework called DIVA to automate the discovery and download of official and non-official images on Docker Hub and the analysis of security vulnerabilities and their propagation.

These works relate to security, but none involves the automated transformation of Dockerfiles. Our study, however, is based on the systematic transformation of an application Dockerfile to give observability and fault injection capabilities to the application without having to change the code. There are studies that explore the theme of automated transformation of DevOps artifacts. Henkel et al. \cite{henkelshipwright} introduced a tool that semi-automatically repairs broken Dockerfile builds. Gallaba and Keheliya \cite{gallaba2018use} presented tools called Hansel and Gretel that detect and automatically correct anti-patterns in Travis configuration files. Wettinger et al. \cite{wettinger2016streamlining} proposed a framework for the automated transformation of DevOps artifacts, specifically those for Chef and Juju, into TOSCA artifacts, which is a standard gaining popularity in the cloud computing community. To our knowledge, automated transformation of Dockerfiles for the purpose of augmenting applications with observability and fault injection, as we do with POBS, is an original concept that has not been researched before.

\subsection{Observability \& Monitoring}

Monitoring a distributed system is a challenging enterprise because of the potentially large number of its constituent components. This problem has been highlighted by Mace et al. \cite{Mace:2018:PTD:3297862.3208104} who implemented a framework called Pivot Tracing which combines dynamic instrumentation of components with causal tracing techniques to collect metrics that span across component and machine boundaries.

Picoreti et al. \cite{8511977} advocated the use of multilevel observability which would include application as well as infrastructure monitoring for the automatic orchestration of microservices deployed on the cloud.

Observability becomes especially significant for chaos engineering and fault injection in order to monitor the state of the system under test before, during, and after an experiment. Several tools such as Grafana \cite{grafana}, Prometheus \cite{prometheus}, DataDog \cite{datadog}, DynaTrace \cite{dynatrace}, cAdvisor \cite{cadvisor}, and others \cite{splunk} \cite{papertrail}, allow the management, visualization, and analysis of monitored data. Bug tech companies like Google and Amazon also have their own observability tools such as Operations (formerly Stackdriver) \cite{Stackdriver} and Amazon CloudWatch \cite{CloudWatch},  both of them require the application to be deployed on their cloud infrastructure. \autoref{tab:related-work-comparison} presents a comparison of POBS against them 4 different dimensions: applied platform, the level of observability, augmented base images, and whether extensible for research. It shows that POBS has unique features compared to those professional tools. First, POBS does not require specific infrastructure platforms. Second, it is the first and only tool that uses Docker image augmentation for observability improvement. Finally, POBS' source code and data being publicly available, it is usable and extensible for future research, contrary to the proprietary systems of Google and Amazon.

In our approach, we improve the observability of an application by generating an augmented base image which includes an observability module and a fault injection module.
This augmented base image then replaces the original base image in the application's Dockerfile. Monitoring is thus available out-of-the-box with virtually no separate tooling or other changes to the application code.


\begin{table}
\centering
\caption{Comparison of existing tools with POBS}\label{tab:related-work-comparison}
\scriptsize
\begin{tabularx}{\columnwidth}{lrrR}
\toprule
& Google & Amazon & \\
Dimension&  Operations& CloudWatch& \textbf{POBS}\\
\midrule
Platform& Google Cloud& AWS& \textbf{Any platform that supports Docker}\\
Observability& Infra, OS& Infra, OS & \textbf{OS, JVM}\\
Augmented base images& No& No& \textbf{Yes}\\
Extensible for research& No& No& \textbf{Yes}\\
\bottomrule
\end{tabularx}
\end{table}

\subsection{Fault Injection}

Previous techniques and tools to perform fault injection on distributed systems include Pumba \cite{Chaos_Testing_Docker}, ChaosCat \cite{ChaosCat}, and Netflix's ChaosMonkey \cite{ChaosMonkey}.

Fault injection tools have also been implemented specifically for microservice-based applications. Simonsson et al. \cite{Simonsson2019ObservabilityAC} implemented a fault injection and monitoring tool for microservice-based applications that use Docker. This tool analyzes the behavior of the application with respect to fault injection on system calls. 
Hiorhiadi et al. \cite{7536505} implemented Gremlin, a tool that can be integrated into the production environment and inject network-level failures by disrupting communication between microservices of an application.

A fault injection tool called CloudVal \cite{5958218} was employed to evaluate the performance and reliability of cloud environments based on KVM and Xen hypervisors with respect to various fault models. 

The fault injection module in POBS instruments Java bytecode to inject exceptions. This module, in conjunction with our observability module, allows developers to assess the resilience of their application at a very low cost.


\section{Conclusion}
In this paper, we have presented POBS, a novel approach that automatically improves the observability of Dockerized Java applications. By evaluating POBS on open-source Docker applications collected from GitHub, we have shown that POBS automatically improves the observability of 72/\nbBaseImages base images. To our knowledge, this is is the first time that automatic improvement of Docker applications is studied.

As future work, we will set up a multi-version runtime for Dockerized applications using POBS for behavior comparison. With our industrial partners, we will integrate POBS in a CI/CD pipeline that enables developers to have advanced resilience engineering with zero manual effort for each new version of their application \cite{arora:replay_without_recording_of_production_bugs}.

\bibliographystyle{plain}
\balance
\bibliography{references}

\end{document}